\documentclass[journal]{IEEEtran}
\usepackage{comment}
\usepackage{cite}
\usepackage{amsmath, amssymb, amsfonts}
\usepackage{algorithmic}
\usepackage{graphicx}
\usepackage{textcomp}
\usepackage{xcolor}
\usepackage{comment}
\usepackage{multirow}
\usepackage{url}
\usepackage{makecell}
\usepackage{soul}
\usepackage{float}
\usepackage{booktabs}

\usepackage{threeparttable} %

\ifCLASSINFOpdf
\else

\fi

\begin{document}

\title{A Time-Domain Dual-Edge Asynchronous Pipelined SAR ADC Featuring Reset-Free Quantization at Multi-GS/s}

\author{Richard~Zeng,~\IEEEmembership{Graduate~Student~Member,~IEEE},                       Anthony~Chan~Carusone,~\IEEEmembership{Fellow,~IEEE},\\
       and~Xilin~Liu,~\IEEEmembership{Senior~Member,~IEEE}
        
\thanks{This work is supported by the Natural Sciences and Engineering Research Council of Canada (NSERC) Alliance Grant, University of Toronto and Alphawave Semi.}

\thanks{Richard Zeng, and Xilin Liu are with the Edward S. Rogers Sr. Department of Electrical and Computer Engineering, University of Toronto, Toronto, ON M5G 2A2, Canada (e-mail: xilinliu@ece.utoronto.ca).}

\thanks{Anthony~Chan~Carusone is with the Edward S. Rogers Sr. Department of Electrical and Computer Engineering, University of Toronto, Toronto, ON M5G 2A2, Canada and AlphaWave Semi, Toronto, ON M5J 2M4, Canada.}
}

\markboth{This work has been submitted to IEEE}
{Shell \MakeLowercase{\textit{et al.}}: Bare Demo of IEEEtran.cls for IEEE Journals}

\maketitle

\begin{abstract}
Time-domain ADCs are attractive for high-speed wireline receivers, as time resolution scales favorably with advanced CMOS technologies, enabling multi-GS/s single-channel sampling rates. However, conventional time-domain ADCs require explicit reset of voltage-to-time and time-domain signal paths between samples, introducing dead time that fundamentally limits resolution, speed, and energy efficiency. This paper introduces a dual-edge reset-free quantization concept for asynchronous pipelined SAR time-domain ADCs, in which both rising and falling signal edges are exploited to enable reset-free quantization within a single conversion period. By eliminating explicit reset phases, the proposed approach expands the effective conversion window and relaxes the resolution–speed tradeoff at high sampling rates. An 8-bit dual-edge asynchronous pipelined SAR time-domain ADC is implemented in 22-nm FD-SOI, incorporating a linearity-compensated dual-edge voltage-to-time converter and a dual-edge time-to-digital converter with independently tunable rising- and falling-edge delays. The prototype occupies a core area of 0.0089~mm$\mathbf{^2}$ and achieves continuous single-channel operation at 3.5 GS/s, with architectural scalability demonstrated through intermittent operation at 10.5 GS/s and higher. At 3.5 GS/s, the ADC achieves 21.6~dB SNDR and 32.2~dB SFDR. The measured performance is primarily limited by identifiable implementation-level factors rather than by architectural constraints, demonstrating the feasibility of dual-edge reset-free quantization for high-speed time-domain ADCs.
\end{abstract}


\IEEEpeerreviewmaketitle

\section{Introduction}

    \begin{figure}[!ht]
        \centering
        \includegraphics[width=1\linewidth]{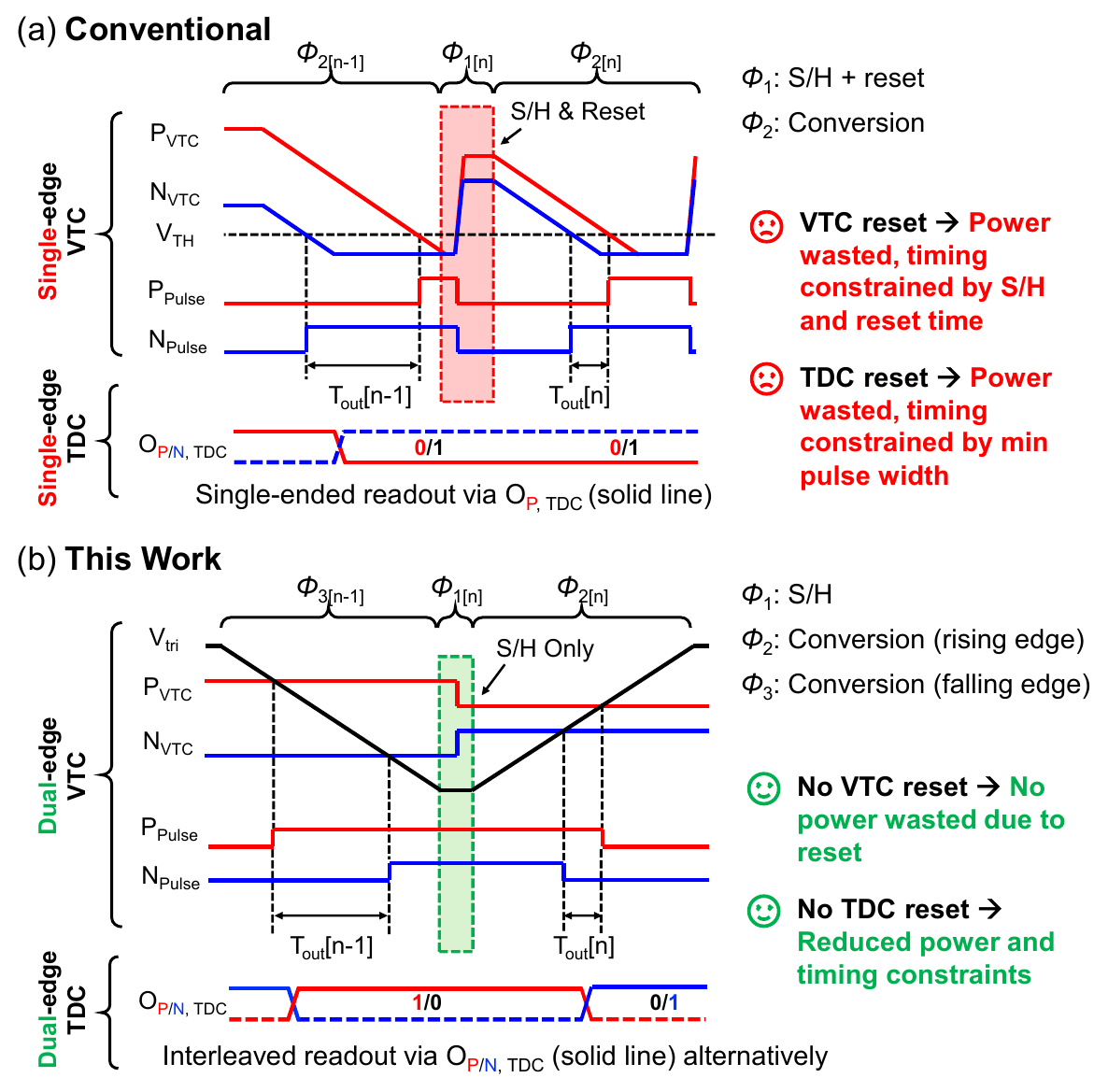}
        \caption{Comparison between (a) conventional single-edge and (b) the proposed dual-edge quantization scheme. }
        \label{fig:archi_cmp}
    \end{figure}

The rapidly increasing data rate demands of high-performance computing, artificial intelligence, and 5G/6G infrastructure require wireline links capable of hundreds of gigabits per second and beyond. ADC-based receivers have become essential for enabling flexible digital equalization and robust system operation. Meeting these requirements necessitates pushing ADC sampling rates to new limits, which places significant design challenges~\cite{I25_TUT_HS_ADC}. Time interleaving (TI) is commonly used to boost the effective sampling rate, however, large interleaving factors complicate timing skew control and require large buffer arrays and clock trees, leading to increased variation sensitivity and power penalties. These challenges drive the need for new ADC architectures and design techniques that achieve high single-channel sampling rates while maintaining ultra compact silicon area.

Time-domain ADCs have emerged as an attractive solution relative to conventional voltage domain architectures~\cite{SSCM25_whitcombe_TD_VD}. A time-domain ADC performs quantization by first converting the sampled input voltage into a time interval using a voltage-to-time converter (VTC), followed by digital quantization in the time domain using a time-to-digital converter (TDC). By operating primarily on time rather than voltage amplitude, these architectures benefit from the continued improvement in delay resolution enabled by scaled CMOS technologies. Furthermore, benefiting from technology scaling, time-domain ADCs can achieve higher single-channel sampling rates, thus requiring fewer time-interleaved channels for target overall sampling rates and improving area efficiencies~\cite{liu_10gss_2022}.

However, as sampling rates extend into the multi-GS/s regime, time-domain ADCs encounter a fundamental resolution versus speed tradeoff due to shrinking quantization windows and increased sensitivity to jitter accumulation along delay elements~\cite{I25_EE3_future_analog}. In feedforward time-domain SAR ADCs, the achievable sampling rate is further constrained by timing overhead associated with pulse generation, propagation, and reset. The requirement for time-to-digital conversion is formalized by Chen \textit{et al.} in the following timing constraint~\cite{chen_178_2023}:
        \begin{equation} \label{eq:ts_min}
            T_\mathrm{S} \geq \frac{T_\mathrm{FS}}{2} + T_\mathrm{M}
        \end{equation}        
where $T_\mathrm{S}$ is the quantization period, $T_\mathrm{FS}$ is the full-scale peak-to-peak time interval, and $T_\mathrm{M}$ is the minimum pulse width within the TDC signal chain to ensure pulse propagation. In conventional designs, $T_\mathrm{M}$ includes time required to reset between consecutive samples. As sampling rates increase, this reset-induced margin increasingly dominates the available conversion window, fundamentally limiting achievable single-channel speed and resolution while consuming dynamic power that does not contribute to quantization. Despite significant advances in time-domain ADCs, most existing designs implicitly accept this limitation by relying on single-edge quantization, in which only one transition of a time pulse carries information while the remainder of the waveform is devoted to reset and reinitialization. As a result, a substantial portion of each conversion period is underutilized, particularly at high sampling rates.

In this work, we introduce a dual-edge quantization concept for asynchronous pipelined SAR time-domain ADCs, in which both rising and falling signal edges are exploited to perform reset-free quantization, featuring a dual-edge VTC followed by a dual-edge TDC, as illustrated in Fig.~\ref{fig:archi_cmp}. By eliminating explicit reset phases, the proposed approach expands the effective conversion window at a given sampling rate, relaxing the fundamental resolution–speed tradeoff imposed by reset overhead while reducing dynamic power associated with pulse resetting. The dual-edge concept is realized within an asynchronous pipelined SAR framework, enabling reuse of a compact delay chain for both edge polarities without requiring a global high-speed clock. 

To validate the proposed concept, an 8-bit dual-edge asynchronous pipelined SAR time-domain ADC is implemented in 22-nm FD-SOI. The prototype incorporates a linearity-compensated dual-edge VTC and a dual-edge TDC with decoupled delay units that independently tune rising- and falling-edge propagation delays. 
Measured results demonstrate continuous single-channel operation at 3.5 GS/s, with architectural scalability explored through intermittent operation at 10.5 GS/s and higher. While the measured performance is limited by identifiable implementation-level constraints, the results experimentally demonstrate the feasibility and scalability of dual-edge reset-free quantization as a new architectural degree of freedom for high-speed time-domain ADCs.

The contributions of this work are summarized below:
\begin{itemize}
    \item A dual-edge quantization concept is introduced for time-domain asynchronous pipelined SAR ADCs, enabling reset-free quantization by exploiting both rising and falling signal edges within a single conversion period.
    \item Within the framework of this innovative design concept, we develop novel circuit implementations of a dual-edge VTC with linearity compensation techniques and a dual-edge TDC featuring dual-edge time comparators and dual-edge decoupled delay units. The reset-free operation reduces the power consumption of the TDC delay chain by up to 40\%.
    \item We present a prototype of an 8-bit asynchronous pipelined SAR ADC fabricated in 22-nm FD-SOI, along with measured results that demonstrate the potential of the architecture and provide insight into practical implementation tradeoffs.
\end{itemize}

The remainder of this paper is structured as follows. Section~\ref{sec:background} reviews essential time-domain ADC design concepts and prior art, Section~\ref{sec:ckt} introduces the proposed dual-edge asynchronous pipelined SAR ADC and describes the circuit implementation details. Section~\ref{sec:meas} presents measurement results and discussion, and Section~\ref{sec:concl} concludes the paper.

\section{Prior Art and Motivation} \label{sec:background}

\subsection{VTC: Linearity and Time Utilization}\label{review:vtc}

VTCs convert sampled voltage into time intervals, typically by charging or discharging capacitive nodes until a threshold crossing occurs~\cite{zhu_time-domain_nodate}. VTC architectures are commonly classified as variable-slope or constant-slope designs. Variable-slope VTCs control the charging or discharging current using the sampled voltage through current sources, offering high-speed operation but suffering from nonlinearity and limited output range. Constant-slope VTCs improve linearity by separating voltage sampling from time generation, at the cost of additional control phases and increased latency.

Several techniques have been proposed to improve VTC linearity and time utilization. Yonar \textit{et al.} proposed a bipolar VTC that removes part of the unused conversion interval by charging and discharging positive and negative sampled voltages separately~\cite{yonar_8b_2023-1, yonar_8b_2023}, but is limited to generating single-sided enable pulses and face challenges near zero-time intervals. Zhao \textit{et al.} introduced a cascaded linearization technique that extends the linear range by combining stages with opposing nonlinearities~\cite{zhao_171_2023}, at the cost of settling latency. Folding VTCs aim to alleviate the time utilization issue by dividing the input voltage range into $2^\mathrm{M}$ segments, each with an output between $2^\mathrm{N}/2^\mathrm{M+1} T_\mathrm{LSB}$. This can reduce the time interval required to quantize each sample, but typically requires multiple VTCs and/or TDCs compared to a single linear VTC, resulting in increased architectural complexity~\cite{jssc_oh_td_flash_2019}.

Several recent designs attempt to reduce the number of VTCs and TDCs used. Yi \textit{et al.} reduced the number of TDCs to one by inserting a multiplexer between VTC and TDC~\cite{yi_4-gss_2021}. Liu \textit{et al.} further reduced the number of VTCs to one by making $\Delta V_\mathrm{th}$ of the VTC adaptive~\cite{I24_liu_td_adc_2025}. Despite reducing calibration overhead due to the reduction of VTC and TDC units, they involve more complex selection logic and circuitry. 

Despite these advances, most existing VTCs share a common drawback: an initial conversion dead time $T_\mathrm{dead}$ during which the voltage ramp is initialized or reset before meaningful time encoding begins. This dead time reduces the effective conversion window available for TDC quantization and becomes increasingly detrimental as sampling rates scale into the multi-GS/s regime.

\subsection{TDC: Throughput Limitations}

TDCs quantize time intervals into digital codes and exhibit tradeoffs among resolution, speed, power, and area~\cite{henzler_10_tdc_basic}. Flash TDCs provide minimal latency but scale poorly in power and area for higher resolutions. Vernier TDCs can achieve high resolution with sub-gate delays but incur long conversion times and high power. Gated ring oscillator TDCs offer a wide dynamic range but require long idle times and are sensitive to jitter accumulation. Pulse-shrinking TDCs can achieve a high dynamic range, but they suffer from high power consumption and require a minimum input time to generate the pulse.

Time amplifiers (TAs) have been introduced to amplify the time interval and improve sensitivity. However, the implementations of TAs suffer from practical design challenges. For example, inverter-based cross-coupled TA suffers from degraded linearity with one-point calibration~\cite{lee_1_2010}, and SR-latch TA is prone to metastability issues when the time input is close to zero~\cite{oulmane_cmos_2004, lee_9_2008}. While linear ramp TA offers an extended input range, it fails to operate effectively near zero time input~\cite{kwon_analysis_2014}. Other TA topologies such as DLL-based and replica-based TAs do not support sub-gate delay resolution and require multiple delay cells, which result in increased power consumption~\cite{nakura_time_2009, kim_7_2013}. 

For wireline receivers requiring high sampling rates and moderate resolution, pipelined SAR TDCs provide an effective tradeoff. Unlike voltage-domain SAR ADCs, time-domain SAR TDCs perform successive approximation by introducing controlled delays to the positive and negative pulses, thereby adjusting their time difference. These architectures are generally classified as feedback-based or feedforward-based. In feedback-based SAR TDCs, the sampling rate is limited because each conversion must complete before the next sample begins~\cite{jssc_09_mantyniemi_sar_tdc}. In contrast, feedforward-based SAR TDCs employ multiple sub-TDCs that each resolve only a portion of the conversion, enabling significantly higher sampling rates \cite{jssc_12_chung_10b_sar_tdc}.
                
Recent SAR TDC research has focused on achieving throughput at multi-GS/s speeds. Delay-tracking pipelined SAR TDCs generate delayed clocks that track the input within each unit SAR cell~\cite{liu_10gss_2022}, achieving 5 GS/s single-channel operation in 14-nm CMOS, but requires calibration for stage delay tracking and suffers from jitter accumulation at LSB stages. Asynchronous pipelined SAR TDCs eliminate external control clocks by generating asynchronous clocks from input pulse combinatorial logic~\cite{chen_178_2023}, reaching 10 GS/s in 28-nm CMOS. However, its reliance on full pulse reset between samples introduces timing overhead and unnecessary dynamic power consumption, ultimately limiting throughput.

\subsection{Limitations of Single-Edge Quantization and Motivation for Dual-Edge Operation}

A common theme across most existing time-domain ADC architectures is the reliance on single-edge quantization. In existing designs, only one transition of a time pulse encodes information, while the remaining portion of the waveform is devoted to reset and reinitialization. This underutilization of the available waveform directly reduces effective conversion efficiency and exacerbates the resolution–speed tradeoff imposed by reset-induced timing overhead.

Dual-edge detection has been explored in feedback-based TDCs to improve power efficiency by utilizing both rising and falling edges~\cite{siddi_double_edge_fb}. However, the feedback nature of these designs imposes critical-path latency constraints that limit achievable sampling rates, and reported implementations have been restricted to sub-GS/s operation. The application of dual-edge concepts to feedforward, asynchronous pipelined SAR architectures operating at multi-GS/s sampling rates has remained unexplored.

The preceding review highlights two fundamental inefficiencies in high-speed time-domain ADCs: conversion dead time in VTCs and reset-induced timing overhead in TDC signal paths. Both limitations stem from single-edge quantization and become increasingly dominant as sampling rates scale upward. Motivated by these observations, this work explores dual-edge quantization as a means to eliminate explicit reset phases and maximize time utilization within each conversion period. By exploiting both rising and falling edges for quantization without explicit reset, the effective conversion window is expanded without extending the sampling period. Given a certain sampling rate, the dual-edge quantization permits a relaxed $T_\mathrm{LSB}$. When combined with an asynchronous pipelined SAR architecture, dual-edge operation enables reuse of a compact delay chain for both edge polarities, reducing dynamic power consumption while alleviating the fundamental resolution–speed tradeoff imposed by reset overhead.

\section{Circuit Implementation} \label{sec:ckt}

\subsection{Architecture Overview}

\begin{figure}[!ht]
    \centering
    \includegraphics[width=1\linewidth]{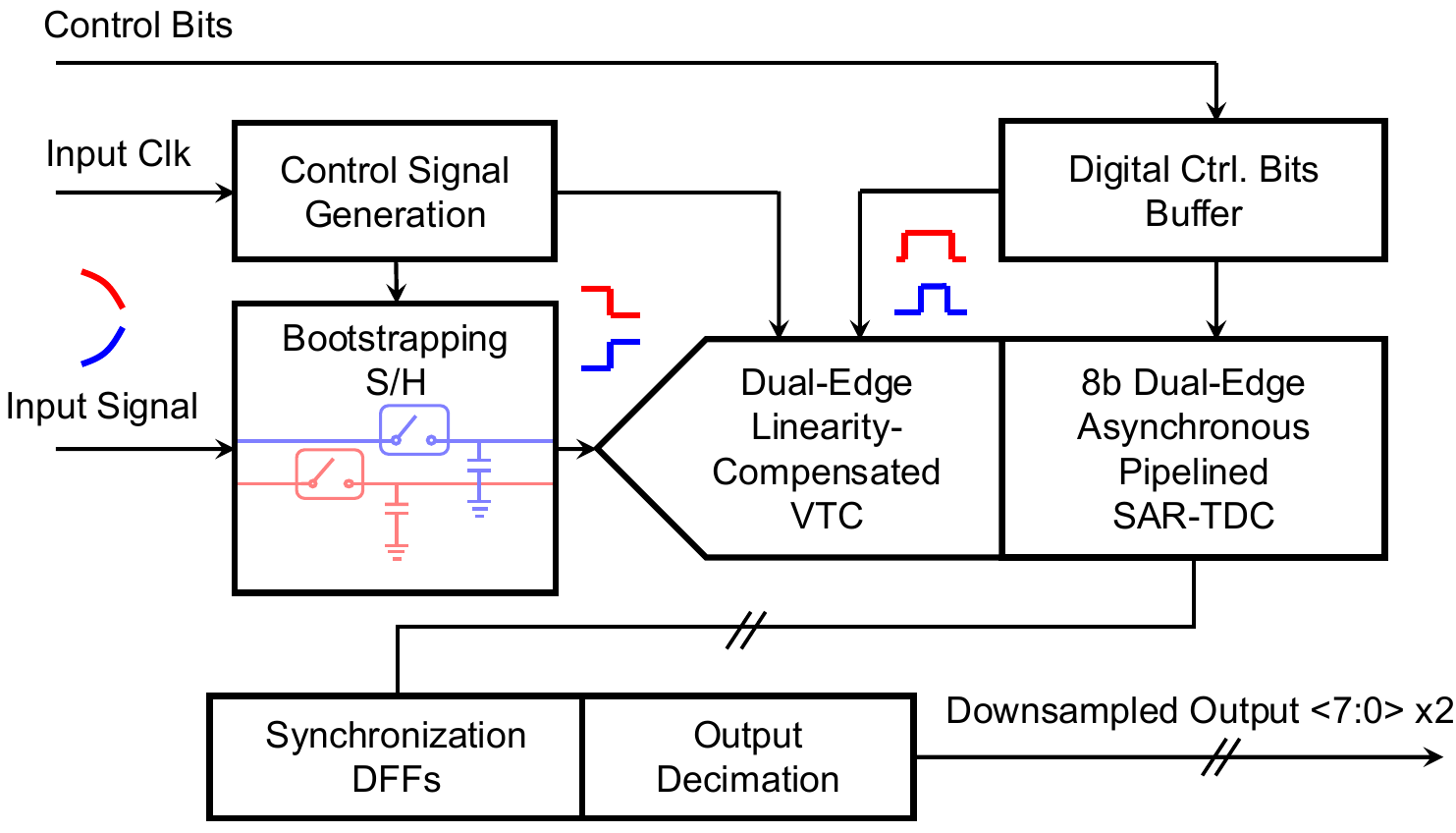}
    \caption{A high-level block diagram of the proposed time-domain dual-edge asynchronous pipelined SAR ADC.}
    \label{fig:tdc_block_diagram}
\end{figure}

A high-level block diagram of the proposed time-domain dual-edge asynchronous pipelined SAR ADC is shown in Fig.~\ref{fig:tdc_block_diagram}. The architecture consists of five main functional blocks: a control signal generation module, a bootstrapped sample-and-hold (S/H), a dual-edge linearity-compensated VTC, an 8-bit dual-edge asynchronous pipelined SAR TDC, and digital synchronization and output decimation logic. The circuit implementation details of the VTC and TDC are discussed in Section~\ref{ssec:impl_de_vtc} and Section~\ref{ssec:impl_de-tdc}, respectively. 

An input clock drives the control signal generation module, which produces timing signals for the S/H and VTC operations. The asynchronously generated digital outputs are synchronized and decimated before sending off-chip. All configuration and calibration control bits are delivered to the chip through shift registers and buffered locally.

\subsection{Dual-Edge VTC with Linearity Compensation} \label{ssec:impl_de_vtc}

\begin{figure}[!ht]
        \centering
        \includegraphics[width=\linewidth]{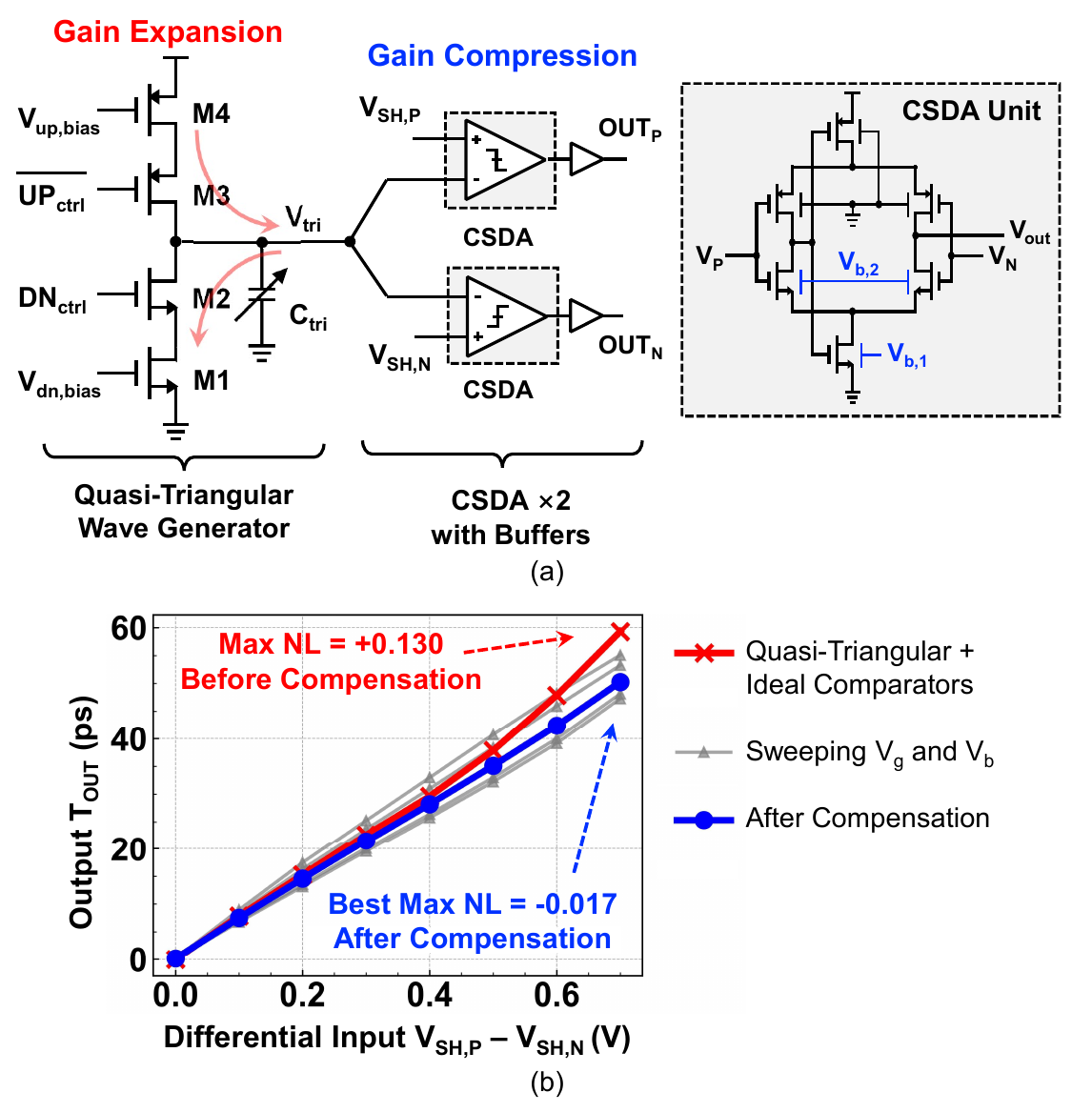}
        \caption{(a) Schematic of the proposed dual-edge VTC and the (b) post-layout simulation results at 12.5~GS/s illustrating the linearity compensation.}
        \label{fig:vtc}
    \end{figure}

As illustrated conceptually in Fig.~\ref{fig:archi_cmp}, the proposed dual-edge VTC operates in three phases: a S/H phase ($\Phi_1$) and two conversion phases that generate rising-edge ($\Phi_2$) and falling-edge ($\Phi_3$) pulses. A simplified schematic of the circuit implementation is shown in Fig.~\ref{fig:vtc}. A quasi-triangular waveform $V_\mathrm{tri}$ is generated by alternately charging and discharging the tunable capacitive node $C_\mathrm{tri}$, controlled by the digital signals $\mathrm{UP_\mathrm{ctrl}}$ and $\mathrm{DN_\mathrm{ctrl}}$, respectively. The rising- and falling-edge pulses, $\mathrm{OUT_P}$ and $\mathrm{OUT_N}$, are generated by comparing $V_\mathrm{tri}$ with the sampled differential voltages $\mathrm{V_{SH,P}}$ and $\mathrm{V_{SH,N}}$, respectively. During the S/H phase, ${V_\mathrm{tri}}$ reaches at either maximum or minimum level, and the subsequent ramp is initiated with minimal dead time. This operation enables compatibility with the asynchronous pipeline and supports high sampling rates. 

A key challenge in VTC design, as reviewed in Section~\ref{review:vtc}, is achieving sufficient linearity without introducing excessive complexity or timing overhead. In the proposed design, simple current sources are used to generate the quasi-triangular waveform $V_{\mathrm{tri}}$, which inherently exhibits an expanding nonlinearity due to channel length modulation and when current source transistors enter triode regions. To compensate for this behavior, a comparator stage based on a complementary self-biased differential amplifier (CSDA) is employed~\cite{bazes_two_1991}, whose transfer characteristic exhibits intrinsic compression nonlinearity. The complementary and self-biased configuration of the CSDA enables high gain, variable effective threshold, and robust operation across PVT variations. By exploiting the back-gate biasing capability of the FD-SOI technology used in this work, the back-gate voltages $V_{\mathrm{b,1}}$ and $V_{\mathrm{b,2}}$ provide fine-grain calibration of the VTC linearity and gain. This is achieved by adjusting the threshold voltage of the NMOS devices and, consequently, the transfer characteristic of the CSDA, allowing the expanding nonlinearity of the waveform generator to be effectively compensated.

The VTC was designed to operate at a target sampling rate of 12.5~GS/s, and Fig.~\ref{fig:vtc}(b) shows the post-layout simulated transfer characteristics of the overall VTC. Prior to compensation, the quasi-triangular waveform generator exhibits a nonlinearity (NL) of 0.130. With compensation provided by the CSDA-based comparator stage, the nonlinearity is reduced to 0.017, corresponding to an approximately $7.6\times$ improvement.
Fig.~\ref{fig:vtc}(b) also demonstrates the tunability of the transfer characteristic through adjustment of the biasing voltages. The resulting linear input range of approximately $\pm 50~$ps is sufficient for operation at 12.5~GS/s. 

The proposed VTC architecture achieves high timing and power efficiency, as no additional phases are used for linearity compensation or calibration. Since a single capacitive node is repeatedly charged and discharged, charge is partially recycled during the discharging phase, resulting in reduced dynamic power compared to conventional single-slope VTC designs. It should be noted that precise matching between the rising and falling slopes of the quasi-triangular waveform is not required. This requirement is relaxed because the subsequent dual-edge TDC stages process rising and falling edges independently, with their quantization paths decoupled.

    \begin{figure*}[!ht]
        \centering
        \includegraphics[width=.8\linewidth]{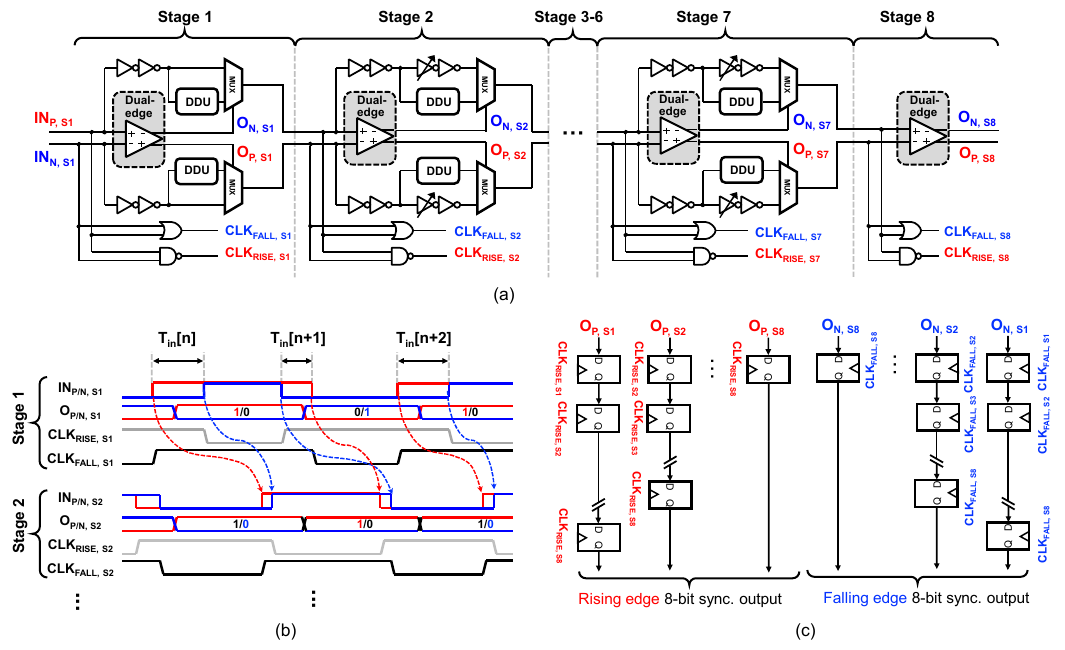}
        \caption{(a) Simplified block diagram of the proposed TDC featuring dual-edge time comparison through asynchronous SAR operation. (b) Simplified timing diagram of the first two TDC stages illustrating the independent rising- and falling-edge decoupled pulse width adjustment. (c) Output bits synchronization scheme with two DFF banks.}
        \label{fig:tdc_combnd_pic}
    \end{figure*}    

\subsection{Dual-Edge Asynchronous Pipelined SAR TDC} \label{ssec:impl_de-tdc}

An 8-bit dual-edge asynchronous pipelined SAR TDC is implemented in this work, as illustrated in Fig.~\ref{fig:tdc_combnd_pic}(a). The TDC independently quantizes the rising- and falling-edge time intervals of the two pulses generated by the dual-edge VTC using a SAR procedure. In each TDC stage, the two pulses are compared by a dual-edge asynchronous time comparator proposed in this work (described in Section~\ref{ssec:impl_detdc_detcmp}), and the resulting comparison outputs ($\mathrm{O_P}$ and $\mathrm{O_N}$) control the multiplexers to select delayed versions of the pulses, as illustrated in Fig.~\ref{fig:tdc_combnd_pic}(b). This operation adjusts the relative timing of the p-side and n-side pulses and progressively reduces the residual time difference, effectively halving the search space at each bit trial until all 8 bits are resolved. Since both rising and falling edges of the two pulses need to be adjusted independently, using only conventional delay units, where the propagation delays of rising and falling transitions are inherently coupled, is not suitable for this work. To address this limitation, a dual-edge decoupled delay unit (DDU) is proposed and is discussed in Section~\ref{ssec:impl_detdc_ddu}. 

Building upon the OR-gate-triggered flip-flop scheme used to capture rising-edge decisions in~\cite{chen_178_2023}, this work extends the approach by incorporating NAND-gate triggering by falling edges. As a result, two separate synchronization banks of D flip-flops are implemented to independently store rising- and falling-edge bit decisions, as illustrated in Fig.~\ref{fig:tdc_combnd_pic}(c). The NAND outputs trigger the capture of the decision bits from the preceding rising edge, whereas the complementary signals handle the falling edge decisions.  All operations of the TDC are performed without an external clock, with combinatorial logic from the input pulses used to capture and synchronize bit decisions across stages. 

With an initial target sampling rate of 12.5~GS/s, the TDC input range is designed to span $\pm$50~ps, corresponding to a nominal time resolution $T_{\mathrm{LSB}}$ of 390.63~fs. Jitter accumulated along the delay chain and observed at the final-stage comparator establishes a lower bound on the achievable $T_{\mathrm{LSB}}$~\cite{chen_8gss_2022}.

\subsubsection{Dual-Edge Asynchronous Time Comparator} \label{ssec:impl_detdc_detcmp}

\begin{figure}[!ht]
    \centering
    \includegraphics[width=\linewidth]{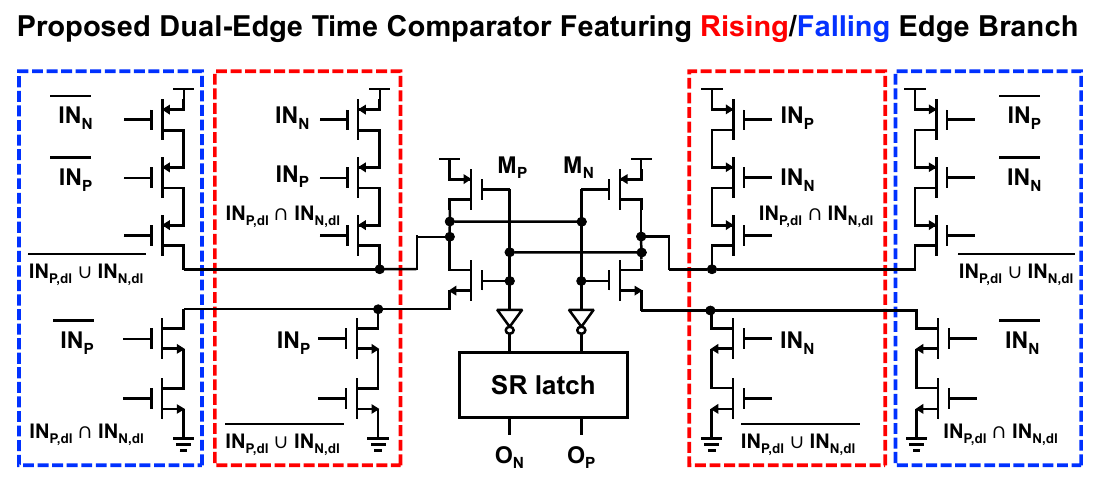}
    \caption{The circuit implementation of the proposed dual-edge time comparator.}
    \label{fig:tdc_de_tcmp}
\end{figure}

To support the dual-edge quantization in the proposed asynchronous TDC, a time comparator capable of comparing both rising and falling transition polarities is proposed for the first time. Fig.~\ref{fig:tdc_de_tcmp} shows the circuit schematic of the comparator design. 
The comparator consists of two dedicated detection branches, one for rising edges and one for falling edges, while sharing a common regeneration and output structure. During rising-edge detection, the rising-edge branch is activated and forms a positive-feedback loop at nodes $\mathrm{M_P}$ and $\mathrm{M_N}$, triggered by the input signals $\mathrm{IN_P}$ and $\mathrm{IN_N}$. The resulting comparison outcome is latched at nodes $\mathrm{O_P}$ and $\mathrm{O_N}$ as a one-bit decision, which is forwarded to the multiplexers and shift registers for stage $\Delta T$ subtraction and bit synchronization. Similarly, during falling-edge detection, the falling-edge branch is activated by $\overline{\mathrm{IN_P}}$ and $\overline{\mathrm{IN_N}}$, while the rising-edge branch remains inactive, allowing the comparator to process the opposite transition polarity using the same circuitry. 

After regeneration, combinatorial logic driven by delayed versions of the input pulses asynchronously resets nodes $\mathrm{M_P}$ and $\mathrm{M_N}$, preparing the comparator for subsequent edge detections without requiring a full pulse reset. By incorporating two complementary detection branches within a shared comparator structure, the proposed design enables reliable detection of both rising and falling edges while reusing the same delay line and comparator circuitry, making it well suited for reset-free dual-edge asynchronous operation.

\subsubsection{Dual-Edge Decoupled Delay Unit (DDU)} \label{ssec:impl_detdc_ddu}

Precise and independent control of the propagation delays associated with both rising and falling signal edges is critical for the TDC proposed in this work. In conventional delay cells, the delays of rising and falling transitions are inherently coupled, which prevents independent control and leads to edge-dependent mismatch accumulation. While prior work has demonstrated single-edge delay manipulation using switch-capacitor circuits and resistive discharge paths for low-frequency applications~\cite{single_edge_dl_lf_22}, such approaches are not suitable at GS/s sampling rates, where tight timing constraints and the effective on-resistance of transistors reduce the need for explicit resistive elements.

\begin{figure}[!ht]
    \centering
    \includegraphics[width=\linewidth]{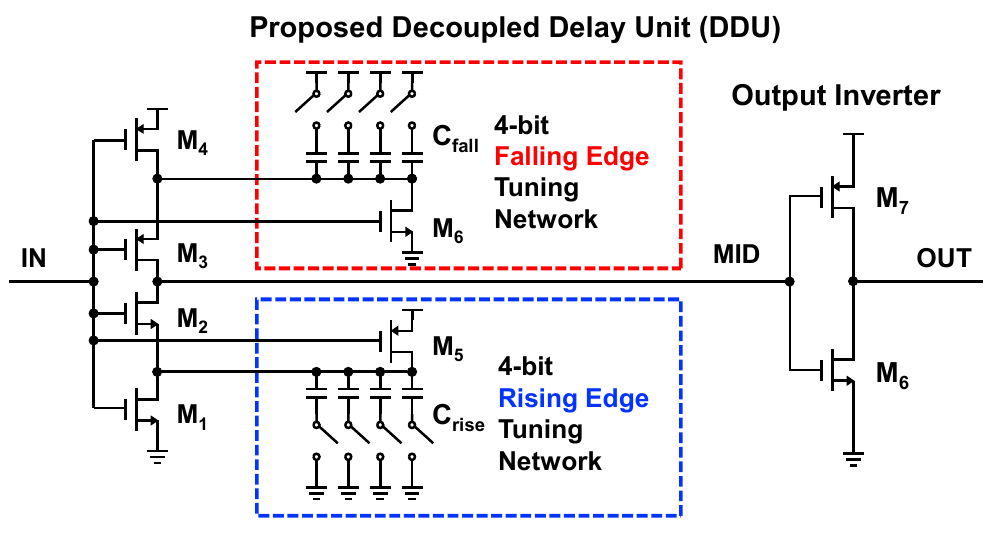}
    \caption{The circuit implementation of the proposed DDU delay unit.}
    \label{fig:tdc_ddu}
\end{figure}

Fig.~\ref{fig:tdc_ddu} shows the schematic of the proposed DDU delay unit, which enables independent tuning of rising- and falling-edge propagation delays through two separate capacitive tuning networks, $C_{\mathrm{rise}}$ and $C_{\mathrm{fall}}$. The operation of the DDU can be illustrated using rising-edge delay adjustment as an example. Prior to the arrival of a rising-edge pulse, $C_{\mathrm{rise}}$ is precharged to the supply voltage. When the rising edge arrives, M5 is turned off and M1 enters saturation, while M2 initially remains off. As the voltage across $C_{\mathrm{rise}}$ decreases, M2 is eventually turned on, allowing node MID to discharge until it crosses the threshold of the output inverter formed by M6 and M7, thereby generating the delayed rising-edge pulse. In this manner, the rising-edge propagation delay is controlled by $C_{\mathrm{rise}}$. An analogous process is employed for falling-edge delay adjustment using the complementary tuning network $C_{\mathrm{fall}}$.

\begin{figure}[!ht]
    \centering
    \includegraphics[width=.9\linewidth]{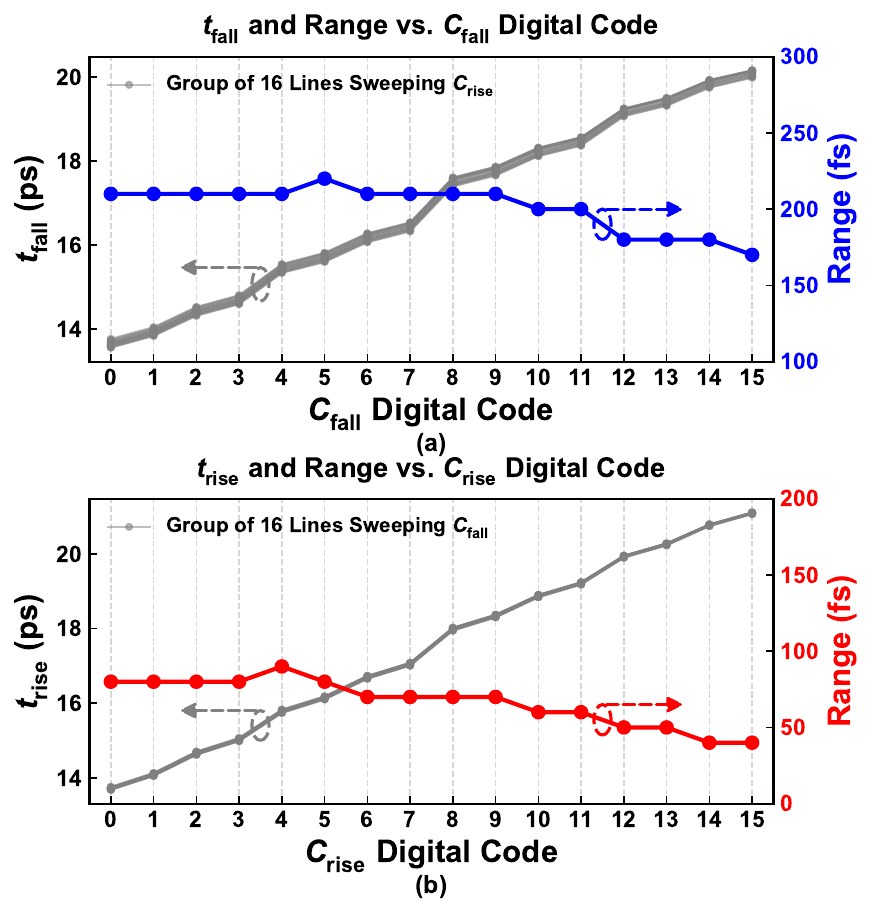}
    \caption{Post-layout simulation of the proposed DDU, including (a) $t_\mathrm{fall}$ and range vs. $C_\mathrm{fall}$ digital code, and (b) $t_\mathrm{rise}$ and range vs. $C_\mathrm{rise}$ digital code. }
    \label{fig:tdc_dcpld_dl_sim}
\end{figure}

Fig.~\ref{fig:tdc_dcpld_dl_sim} shows post-layout simulation results that characterize the rising- and falling-edge delay isolation of the proposed DDU. When sweeping $C_{\mathrm{fall}}$, the resulting variation in the rising-edge delay $t_{\mathrm{rise}}$ is confined within 220~fs, while sweeping $C_{\mathrm{rise}}$ induces less than 90~fs variation in the falling-edge delay $t_{\mathrm{fall}}$. This small coupling is attributed to parasitic capacitances at node MID, which introduce limited cross-coupling. The worst-case coupling is well below one LSB of the designed time resolution and deterministic in nature, therefore it will not limit calibration convergence or overall TDC resolution. These results confirm effective delay decoupling and validate the proposed DDU design.

In the implemented TDC, conventional delay units are combined with the proposed DDUs from stages 2 to 7 to achieve finer $\Delta T$ calibration granularity, while the first stage omits conventional delay units to support a larger initial $\Delta T$, as illustrated in Fig.~\ref{fig:tdc_combnd_pic}(a). Since conventional delay units do not require explicit reset before processing the next sample in dual-edge operation, this approach yields significant power savings. Simulations show that the proposed TDC delay chain achieves up to 40\% power reduction compared to single-edge counterparts.

\section{Measurements and Discussion} \label{sec:meas}

A silicon prototype of the developed 8-bit time-domain dual-edge asynchronous pipelined SAR ADC has been fabricated in GlobalFoundries 22~nm FD-SOI and fully tested on bench. The prototype was initially designed to operate at sampling rates of up to 12.5~GS/s. However, due to specific implementation factors, the fabricated silicon exhibits timing deviations that limit performance. Nevertheless, the measurements demonstrate that dual-edge, reset-free quantization is a viable and scalable architectural approach for high-speed time-domain ADCs, with the observed limitations arising from localized and addressable implementation issues rather than fundamental architectural constraints.

To comprehensively present the measurement results, we first introduce the prototype and measurement setup, followed by validation of the dual-edge quantization and performance characterization. We then discuss the implementation limitations and their root causes in detail, along with mitigation strategies. Finally, we present the power consumption, scaling trends, and a comparative analysis with the state of the art.

\subsection{Prototype and Measurement Setup}
\begin{figure}[!ht]
    \centering
    \includegraphics[width=.95\linewidth]{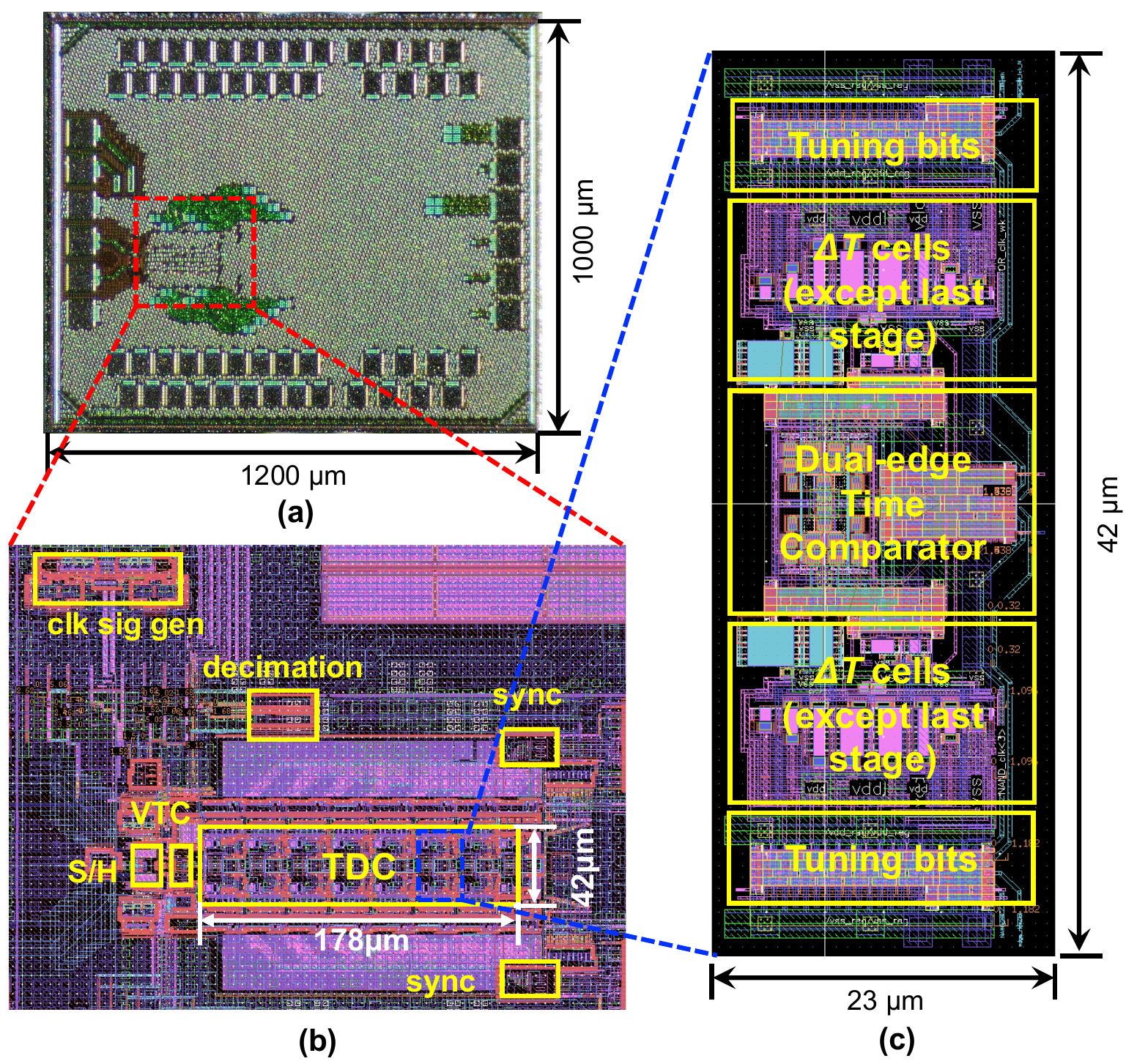}
    \caption{(a) The micrograph of the fabricated chip in 22-nm FD-SOI. (2) Layout of the with key circuit blocks highlighted. (3) Zoomed-in view of one TDC stage.}
    \label{fig:chip_die_photo}
\end{figure}

Fig.~\ref{fig:chip_die_photo} shows a micrograph of the fabricated chip, along with a zoomed-in view of the layout highlighting the key building blocks. 
One TDC stage is highlighted, showing a symmetrical floorplan centered on the dual-edge time comparator. This structure is repeated across stages 1 to 7, with stage 8 slightly different due to the absence of the $\Delta$T cells. Digital logic blocks from Invecas were manually placed and routed to improve achievable speed. Different blocks are isolated via multiple layers of guard rings. The active area of the entire ADC is 0.0089~mm$\mathrm{^2}$, with the TDC core occupying less than 0.0075~mm$\mathrm{^2}$, among the smallest reported for comparable pipelined TDCs.

\begin{figure}[!ht]
    \centering
    \includegraphics[width=\linewidth]{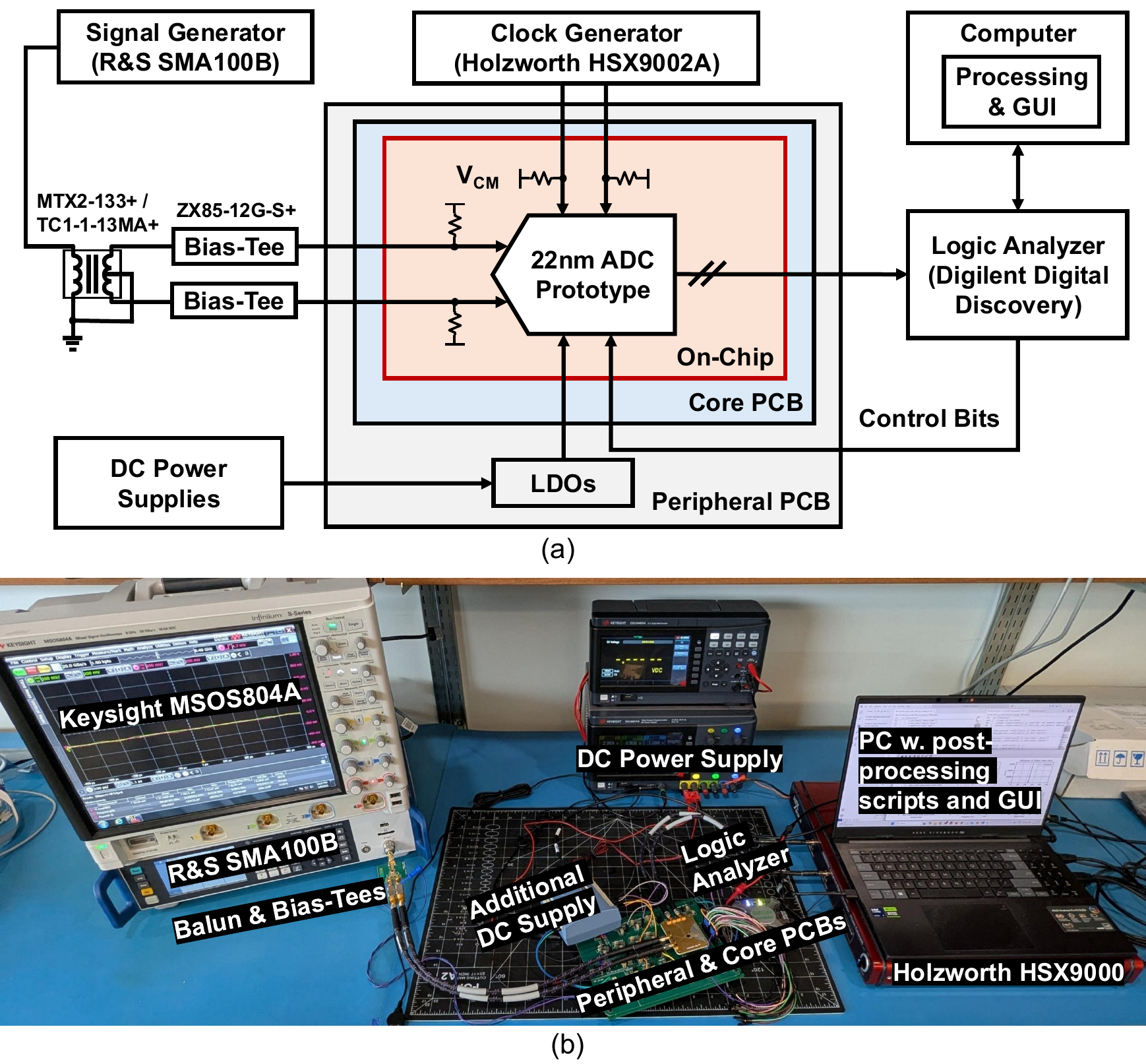} 
    \caption{(a) The block diagram and (b) a photo of the bench measurement setup used for characterizing the ADC.}
    \label{fig:meas_setup_actual_note}
\end{figure}

Fig.~\ref{fig:meas_setup_actual_note} shows the bench measurement setup. Chip-on-Board (CoB) bonding was used with signal pads routing deliberately shortened and aligned. The input signal comes from an R\&S SMA100B and routed through baluns (MTX2-133+ and TC1-1-13MA+, Mini-Circuits) for high- and low-frequency measurements, respectively. A logic analyzer (Digital Discovery, Digilent) is used to program the digital control bits and to capture the decimated output data from the chip, with a tunable decimation factor of up to 525$\times$.

\begin{figure}[!ht]
    \centering
    \includegraphics[width=.7\linewidth]{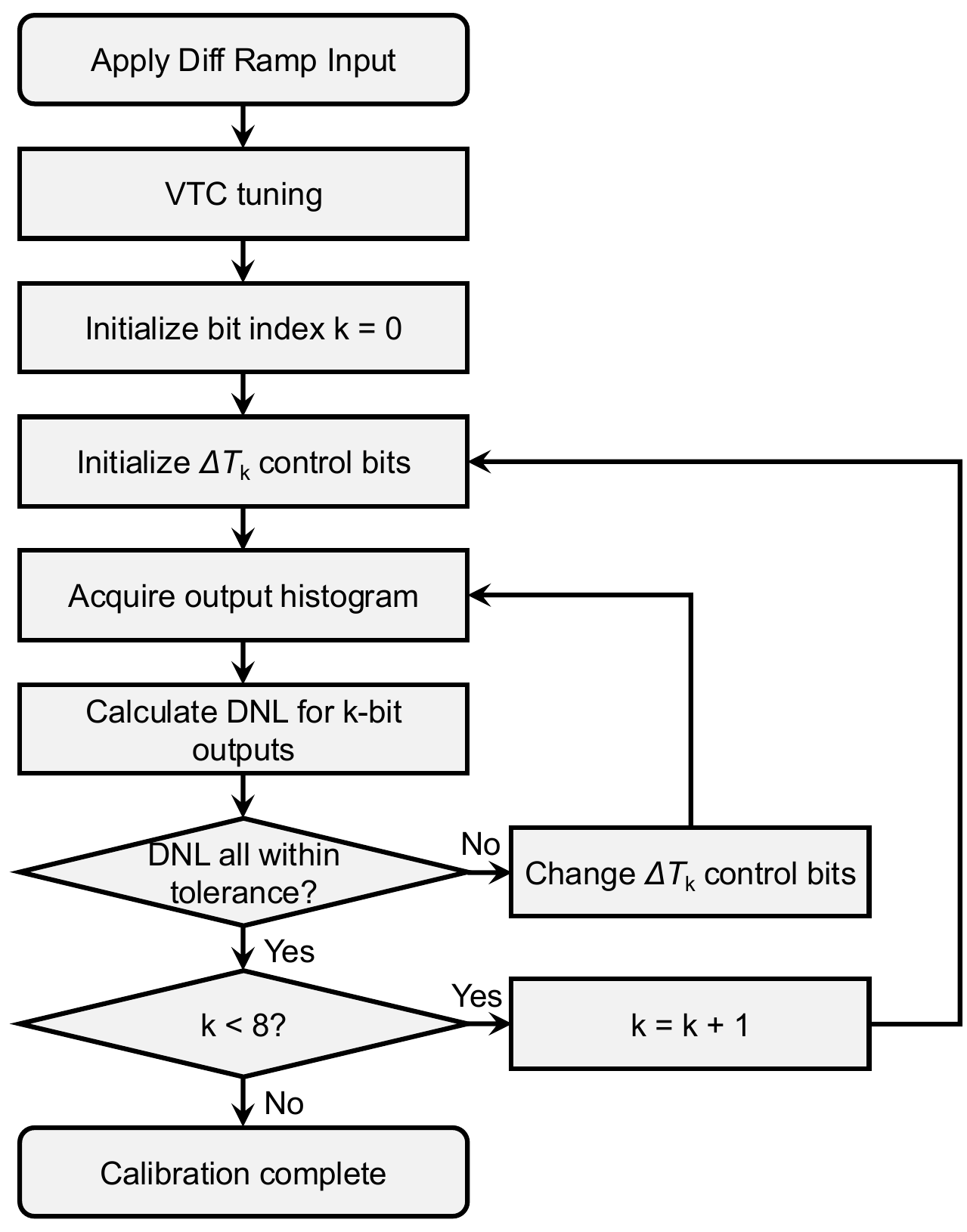}    
    \caption{The flow chart of the developed foreground calibration scheme.}
    \label{fig:meas_cal_flow}
\end{figure}

A foreground calibration procedure is applied to the prototype chip, as illustrated in Fig.~\ref{fig:meas_cal_flow}. Calibration begins by applying a differential ramp input to facilitate VTC tuning. The procedure then iterates through the TDC stages sequentially by bit index, starting from $k=0$. At each step, the delay control bits $\Delta T_\mathrm{k}$ are initialized, the first $(k+1)$ output bits are enabled, and an output histogram is collected to evaluate the differential nonlinearity (DNL). If the measured DNL exceeds a predefined tolerance window, the $\Delta T_\mathrm{k}$ control bits are adjusted and the histogram acquisition is repeated. Once the linearity criterion is satisfied, the bit index $k$ is incremented and the process continues until all bits are calibrated, at which point the foreground calibration is complete.

\begin{figure*}[!ht]
    \centering
    \includegraphics[width=.65\linewidth]{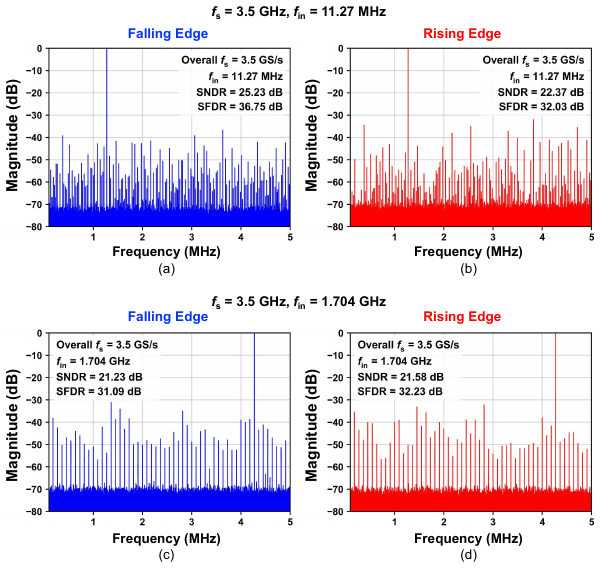}
    \caption{Measured output spectra using a 65,536-point FFT with an output decimation factor of 175: (a) low-frequency input, falling-edge quantization; (b) low-frequency input, rising-edge quantization; (c) high-frequency input, falling-edge quantization; and (d) high-frequency input, rising-edge quantization.}
    \label{fig:meas_LF_HF}
\end{figure*}

\subsection{Validation of Dual-Edge Quantization}

The fundamental operation of the proposed dual-edge quantization scheme is first validated by evaluating the independent performance of the rising- and falling-edge signal paths. Fig.~\ref{fig:meas_LF_HF} shows representative output spectra obtained from the rising-edge and falling-edge quantization paths with both low and high input frequencies at a sampling rate of 3.5~GS/s. Both edges exhibit clear spectral tones and comparable noise floors, confirming that input information is successfully encoded. The similar spectral characteristics observed across the two paths indicate that the dual-edge VTC and comparator circuitry operate as intended, and that no inherent asymmetry prevents effective quantization on either edge. 
 
\subsection{Performance Characterization}

\begin{figure}[!ht]
    \centering
    \includegraphics[width=.85\linewidth]{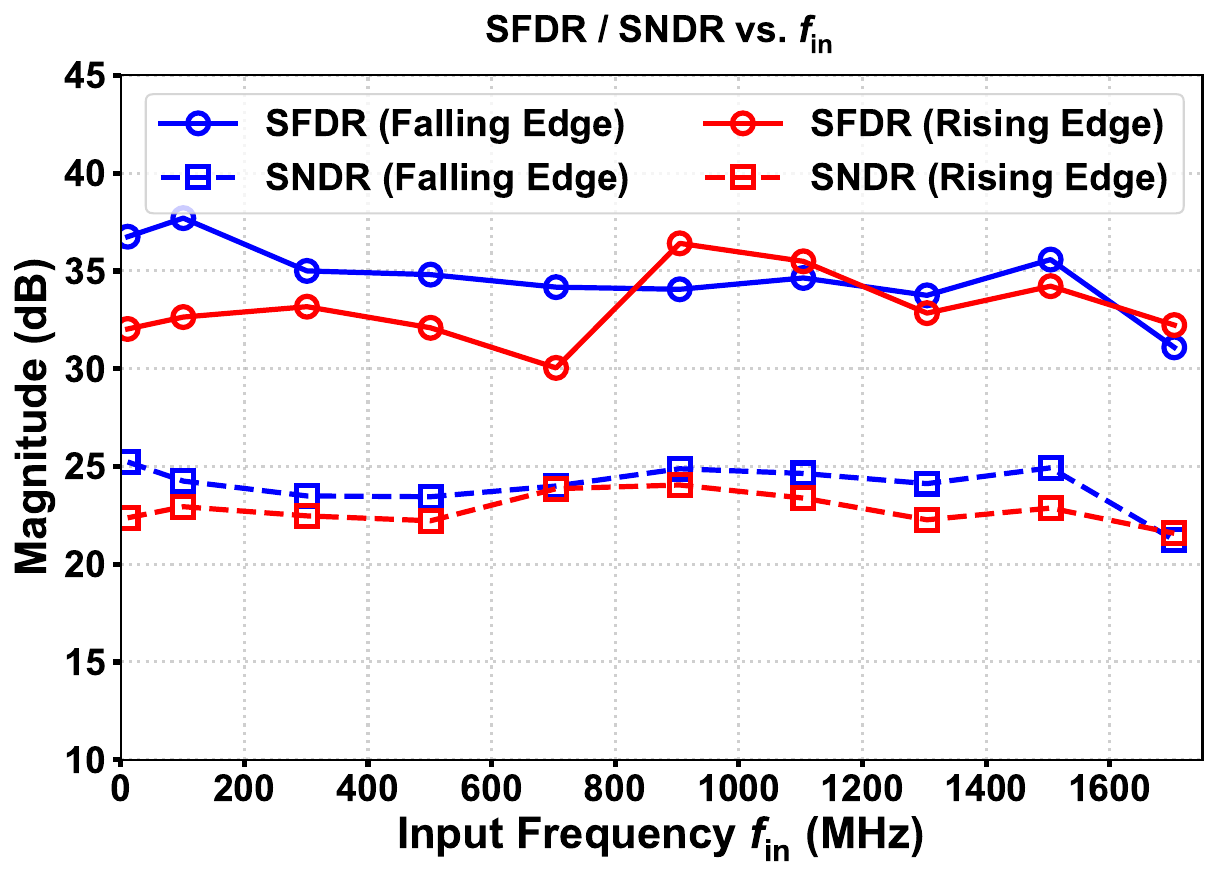}
    \caption{Measured SNDR and SFDR versus input frequencies at 3.5~GS/s.}
    \label{fig:meas_SNDR_SFDR}
\end{figure}

Fig.~\ref{fig:meas_SNDR_SFDR} shows the measured signal-to-noise-and-distortion ratio (SNDR) and spurious-free dynamic range (SFDR) as a function of input frequency at a sampling rate of 3.5~GS/s. Stable operation is observed up to Nyquist frequency for both rising- and falling-edge quantization paths, demonstrating correct timing-domain signal propagation and asynchronous SAR operation across the full input bandwidth. The similar frequency-dependent trends further confirm that dual-edge quantization does not introduce a fundamental bandwidth limitation.

\begin{figure}[!ht]
    \centering
    \includegraphics[width=.8\linewidth, height=6.5cm]{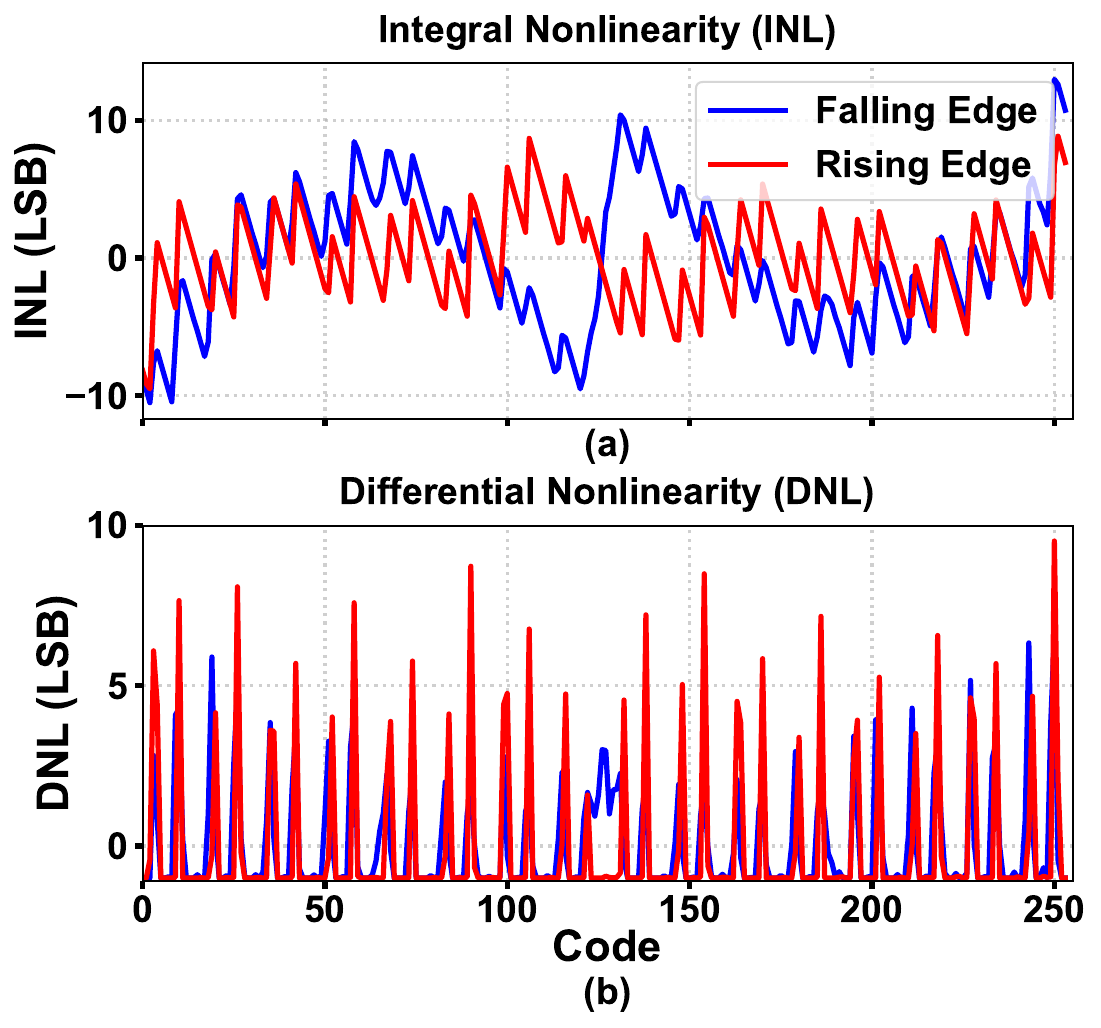}\vspace{-3mm}
    \caption{Measured (a) INL and (b) DNL at 3.5~GS/s.}
    \label{fig:meas_inl_dnl}
\end{figure}

Fig.~\ref{fig:meas_inl_dnl} shows the measured DNL and INL. The DNL exhibits a periodic structure rather than random variation, indicating the presence of deterministic timing mismatch rather than noise-dominated behavior. This periodicity corresponds to specific SAR stages and is consistent with $\Delta$T mismatches in the TDC stages.

The strong correspondence between the DNL curve and the SNDR degradation indicates that the performance is primarily limited by stage-level timing mismatch, rather than by the dual-edge quantization principle itself. Since the mismatch is confined to a small subset of SAR stages, improving the tuning granularity and matching of the affected delay elements is expected to yield a disproportionate improvement in both linearity and SNDR, which will be discussed in Section~\ref{ssec:limitation}.

\begin{figure}
    \centering
    \includegraphics[width=.65\linewidth]{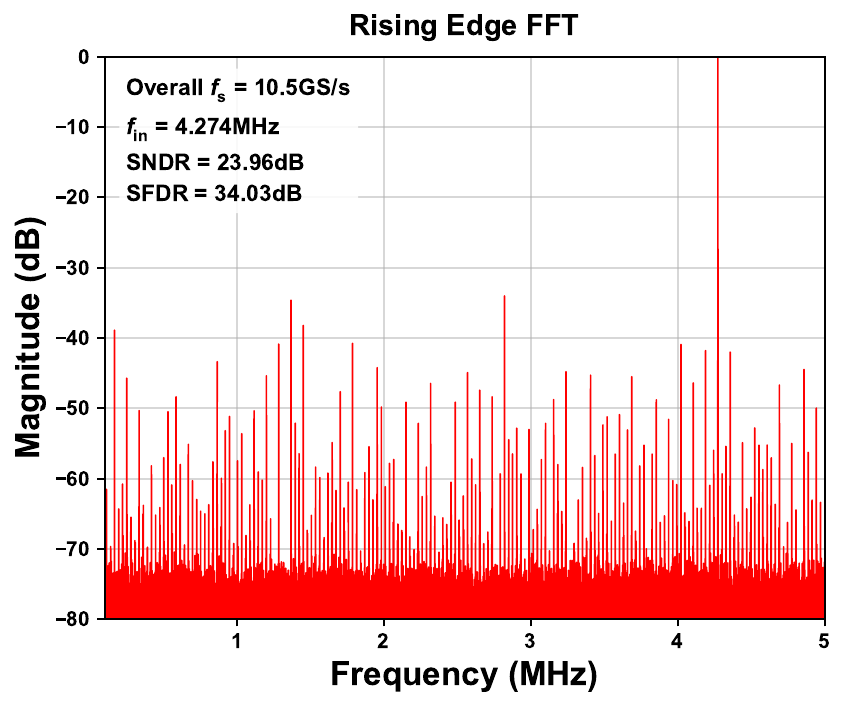}
    \caption{A representative spectrum during intermittent operation at a sampling rate of 10.5~GS/s (decimated by 525).}
    \label{fig:oneshot}
\end{figure}

To evaluate architectural scalability, the ADC was exercised at higher sampling rates. Fig.~\ref{fig:oneshot} shows a representative output spectrum obtained at a sampling rate of 10.5~GS/s. Although the measurement does not represent reliable continuous operation, it demonstrates that the proposed dual-edge asynchronous pipelined SAR architecture does not impose a fundamental speed limitation at these sampling rates. 

\subsection{Prototype Limitations and Mitigation Strategies}\label{ssec:limitation}

The measured performance limitations of the prototype chips can be attributed to a small number of identifiable and localized implementation factors, mainly the packaging parasitics and delay tuning mismatches in the TDC. 

\subsubsection{External Interface and Signal Integrity}
    The measurement results indicate that the current CoB packaging imposes an extrinsic limitation on the prototype's high-frequency linearity, masking the intrinsic performance of the core circuitry. While PCB design minimizes and balances trace lengths, the manual wire-bonding process introduces stochastic variations in bondwire length and curvature. These physical irregularities create parasitic inductance mismatches between the differential input paths, resulting in unavoidable amplitude and phase skew at high frequencies, and consequently degraded harmonic distortion performance. 
    
    This limitation can be mitigated by transitioning to a direct probing setup. Using a probe station with high-precision Ground-Signal-Ground-Signal-Ground (GSGSG) RF probes will bypass the bondwires entirely. Together with de-embedding techniques, this will ensure an amplitude- and phase-balanced differential signal input. Such a configuration is expected to recover signal integrity degraded by packaging asymmetries.

\subsubsection{Quantization Noise and Tuning Granularity}
    The primary bottleneck limiting the SNDR is a systematic non-linearity in the TDC, manifested as periodic missing codes in the measured DNL. Detailed stage-by-stage tuning with low-frequency ramp input indicates that delay mismatch buildup exceeds the calibration capabilities of current delay units. This behavior is likely caused by larger parasitic effects than anticipated from post-layout simulations, resulting in less granular $\Delta T$s relative to the designed $T_{\mathrm{LSB}}$.

\begin{figure}
        \centering
        \includegraphics[width=0.75\linewidth]{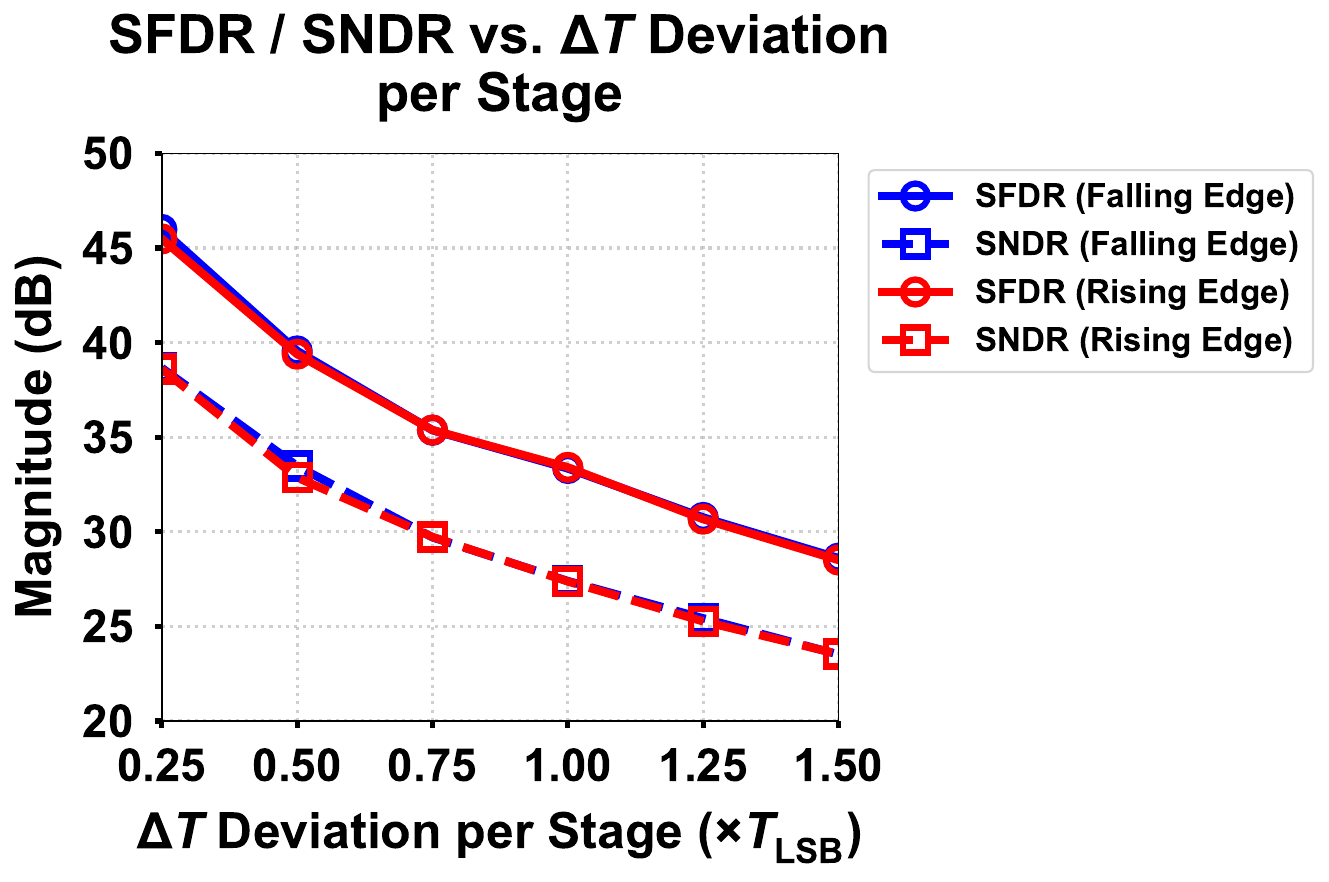}
        \caption{Behavioral simulation of per stage TDC $\Delta T$ deviation vs. SNDR/SFDR plot.}
        \label{fig:xtra_dt_bit_comp}
    \end{figure}

    To validate this hypothesis, a behavioral model was developed to evaluate the impact of $\Delta T$ variation on SNDR and SFDR. The model incorporates stage-dependent $\Delta T$ deviations and includes additive Gaussian jitter to emulate realistic delay uncertainty. Fig.~\ref{fig:xtra_dt_bit_comp} shows the simulation results, which closely match the degradation trends observed in measurement. These results confirm that the observed performance limitation arises from $\Delta T$ variation that cannot be fully corrected through the existing calibration scheme, leading to lower-than-expected linearity performance in the TDC.
    
    This limitation can be addressed by increasing the tuning granularity of the conventional delay units. In the current implementation, the stage $\Delta T$ is realized using a combination of the proposed DDUs and conventional delay cells, providing 8 bits of tunability (4 bits for $\Delta t_{\mathrm{rise}}$ and 4 bits for $\Delta t_{\mathrm{fall}}$) and 2 bits of tunability, respectively. Since adjustments to the conventional delay units affect rising- and falling-edge delays to different extents, modifications to these units often require corresponding updates to both DDU tuning paths. Adding one to two additional tuning bits to the conventional delay units would provide greater degrees of freedom to accurately set the desired $\Delta T$ and is expected to enable SNDR exceeding 35~dB, as indicated by the behavioral simulation results in Fig.~\ref{fig:xtra_dt_bit_comp}.

\begin{table*}[!ht]
\renewcommand{\arraystretch}{1.3}
\caption{Comparative Analysis with the State-of-the-Art High-Speed ADC Designs}
\label{tab:performance_comparison}
\centering
\begin{threeparttable}
\begin{tabular}{|c|cc|cccccccc|}
\hline
                                 & \multicolumn{1}{c|}{\begin{tabular}[c]{@{}c@{}}TCAS I\\ 25 \cite{T125_Jara_pred_SAR}\end{tabular}} & \begin{tabular}[c]{@{}c@{}}JSSC\\ 25 \cite{J25_Chak_7b}\end{tabular} & \multicolumn{1}{c|}{\begin{tabular}[c]{@{}c@{}}JSSC\\ 18 \cite{zhu_2-gss_2018}\end{tabular}} & \multicolumn{1}{c|}{\begin{tabular}[c]{@{}c@{}}JSSC\\ 21 \cite{yi_4-gss_2021}\end{tabular}} & \multicolumn{2}{c|}{\begin{tabular}[c]{@{}c@{}}ISSCC\\ 22 \cite{liu_10gss_2022}\end{tabular}}                     & \multicolumn{1}{c|}{\begin{tabular}[c]{@{}c@{}}ISSCC\\ 25 \cite{I24_liu_td_adc_2025}\end{tabular}} & \multicolumn{1}{c|}{\begin{tabular}[c]{@{}c@{}}JSSC\\ 22 \cite{chen_8gss_2022}\end{tabular}} & \multicolumn{1}{c|}{\begin{tabular}[c]{@{}c@{}}ISSCC\\ 23 \cite{chen_178_2023}\end{tabular}} & \textbf{This Work} \\ \hline
\multirow{3}{*}{Architecture}    & \multicolumn{2}{c|}{Voltage-Domain ADCs}                                                                                     & \multicolumn{8}{c|}{Time-Domain ADCs}                                                                                                                                                                                                                                                                                                                                                                                                                                                           \\ \cline{2-11} 
                                 & \multicolumn{1}{c|}{TI-SAR}                                              & Async. SAR                                        & \multicolumn{1}{c|}{Flash}                                             & \multicolumn{1}{c|}{Pipelined}                                         & \multicolumn{2}{c|}{\begin{tabular}[c]{@{}c@{}}Delay Tracking\\ Pipelined SAR\end{tabular}} & \multicolumn{4}{c|}{Async. Pipelined SAR}                                                                                                                                                                                                       \\ \cline{2-11} 
                                 & \multicolumn{1}{c|}{/}                                                   & /                                                 & \multicolumn{7}{c|}{Single-Edge}                                                                                                                                                                                                                                                                                                                                                                                                                                           & \textbf{Dual-Edge} \\ \hline
Technology (nm)                  & \multicolumn{1}{c|}{28}                                                  & 3                                                 & \multicolumn{1}{c|}{65}                                                & \multicolumn{1}{c|}{65}                                                & \multicolumn{2}{c|}{14}                                                                     & \multicolumn{1}{c|}{28}                                                 & \multicolumn{1}{c|}{28}                                                & \multicolumn{1}{c|}{28}                                                 & 22                 \\ \hline
Supply (V)                       & \multicolumn{1}{c|}{0.8}                                                 & 0.65                                              & \multicolumn{1}{c|}{1}                                                 & \multicolumn{1}{c|}{1}                                                 & \multicolumn{2}{c|}{0.8}                                                                    & \multicolumn{1}{c|}{0.9}                                                & \multicolumn{1}{c|}{0.9}                                               & \multicolumn{1}{c|}{0.9}                                                & 0.9                \\ \hline
Power (mW)                       & \multicolumn{1}{c|}{19.7}                                                & 0.69                                              & \multicolumn{1}{c|}{21}                                                & \multicolumn{1}{c|}{11.3}                                              & \multicolumn{1}{c|}{7.4}                     & \multicolumn{1}{c|}{14.8}                    & \multicolumn{1}{c|}{22.9}                                               & \multicolumn{1}{c|}{85.3}                                              & \multicolumn{1}{c|}{31.7}                                               & 28.3               \\ \hline
Resolution (bit)                 & \multicolumn{1}{c|}{6}                                                   & 7                                                 & \multicolumn{1}{c|}{8}                                                 & \multicolumn{1}{c|}{7}                                                 & \multicolumn{1}{c|}{8}                       & \multicolumn{1}{c|}{8}                       & \multicolumn{1}{c|}{8}                                                  & \multicolumn{1}{c|}{8}                                                 & \multicolumn{1}{c|}{8}                                                  & 8                  \\ \hline
TI Channels No.                  & \multicolumn{1}{c|}{8}                                                   & 1                                                 & \multicolumn{1}{c|}{1}                                                 & \multicolumn{1}{c|}{1}                                                 & \multicolumn{1}{c|}{1}                       & \multicolumn{1}{c|}{2}                       & \multicolumn{1}{c|}{1}                                                  & \multicolumn{1}{c|}{1}                                                 & \multicolumn{1}{c|}{1}                                                  & 1                  \\ \hline
$f_\mathrm{s}$ (GS/s)            & \multicolumn{1}{c|}{10}                                                  & 1.75                                              & \multicolumn{1}{c|}{2}                                                 & \multicolumn{1}{c|}{4}                                                 & \multicolumn{1}{c|}{5}                       & \multicolumn{1}{c|}{10}                      & \multicolumn{1}{c|}{3}                                                  & \multicolumn{1}{c|}{8}                                                 & \multicolumn{1}{c|}{10}                                                 & 3.5\tnote{1}                \\ \hline
$f_\mathrm{s}$ per channel (GS/s) & \multicolumn{1}{c|}{1.25}                                                & 1.75                                              & \multicolumn{1}{c|}{2}                                                 & \multicolumn{1}{c|}{4}                                                 & \multicolumn{1}{c|}{5}                       & \multicolumn{1}{c|}{5}                       & \multicolumn{1}{c|}{3}                                                  & \multicolumn{1}{c|}{8}                                                 & \multicolumn{1}{c|}{10}                                                 & 3.5\tnote{1}                \\ \hline
Area (mm$^2$)                    & \multicolumn{1}{c|}{0.14}                                                & 0.00055                                           & \multicolumn{1}{c|}{0.08}                                              & \multicolumn{1}{c|}{0.011}                                             & \multicolumn{1}{c|}{0.0014}                  & \multicolumn{1}{c|}{0.0029}                  & \multicolumn{1}{c|}{0.032}                                              & \multicolumn{1}{c|}{0.011}                                             & \multicolumn{1}{c|}{0.0091}                                             & 0.0089             \\ \hline
SNDR (dB) @$f_\mathrm{nyq}$      & \multicolumn{1}{c|}{31.2}                                                & 37                                                & \multicolumn{1}{c|}{40.7}                                              & \multicolumn{1}{c|}{34.6}                                              & \multicolumn{1}{c|}{40.8}                    & \multicolumn{1}{c|}{37.2}                    & \multicolumn{1}{c|}{49.3}                                               & \multicolumn{1}{c|}{39.2}                                              & \multicolumn{1}{c|}{36.4}                                               & 21.23\tnote{2}        \\ \hline
SFDR (dB) @$f_\mathrm{nyq}$      & \multicolumn{1}{c|}{44.5}                                                & 46.0                                              & \multicolumn{1}{c|}{48.4}                                              & \multicolumn{1}{c|}{45.7}                                              & \multicolumn{1}{c|}{53.1}                    & \multicolumn{1}{c|}{50.69}                   & \multicolumn{1}{c|}{70.9}                                               & \multicolumn{1}{c|}{56.1}                                              & \multicolumn{1}{c|}{51.7}                                               & 31.09\tnote{2}       \\ \hline
Energy per sample (pJ)    & \multicolumn{1}{c|}{1.97}                                                & 0.39                                              & \multicolumn{1}{c|}{10.5}                                              & \multicolumn{1}{c|}{2.825}                                             & \multicolumn{1}{c|}{1.48}                    & \multicolumn{1}{c|}{1.48}                    & \multicolumn{1}{c|}{7.63}                                               & \multicolumn{1}{c|}{10.66}                                             & \multicolumn{1}{c|}{3.17}                                               & 8.08\tnote{3}               \\ \hline
\end{tabular}
\begin{tablenotes}
      \scriptsize
      \item[1] Continuous operation is demonstrated at 3.5~GS/s; intermittent operation observed up to 10.5~GS/s.
      \item[2] Performance limited by identifiable implementation-level constraints and does not reflect intrinsic architectural efficiency. Worse results between rising/falling edges are reported. \item[3] At 10.5 GS/s intermittent operation, the measured energy per sample reduces to 6.35~pJ/sample. 
    \end{tablenotes}
\end{threeparttable}
\end{table*}


\subsubsection{Time Comparator Metastability}

    The performance of the dual-edge asynchronous time comparator is critical to this design and places stringent requirements on its regeneration speed. In the proposed architecture, the associated switching network introduces additional parasitic capacitance, which degrades the regeneration time constant and increases latency variability. To ensure reliable operation of the dual-edge asynchronous pipelined SAR architecture, the comparator regeneration time must therefore be tightly bounded. Incorporating a metastability resolution circuit that forces a deterministic output after a predefined decision latency enables robust comparison when the input time difference approaches zero.

\subsection{Power Consumption and Scaling Trends}

\begin{figure}[!ht]
    \centering
    \includegraphics[width=\linewidth]{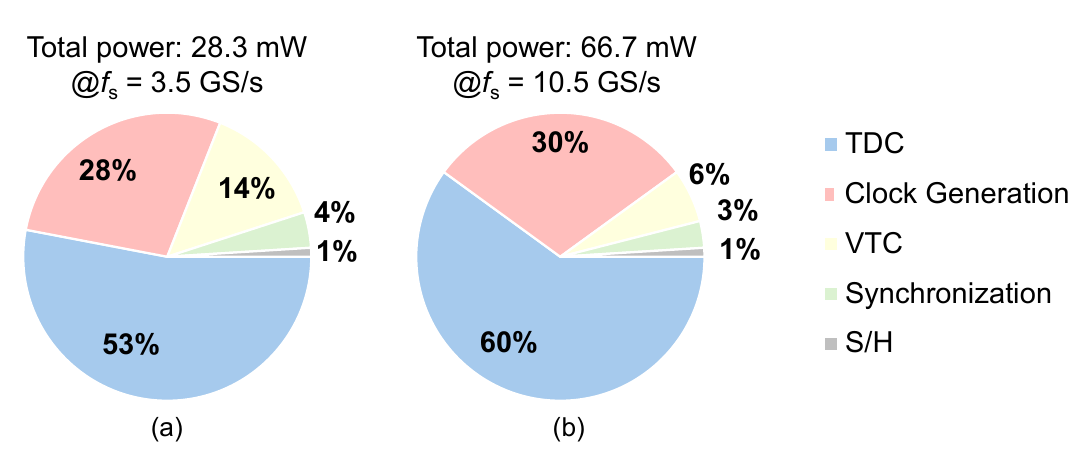}
    \caption{Measured power compositions at (a) 3.5~GS/s and (b) 10.5~GS/s with a supply voltage of 0.9~V. }
    \label{fig:meas_adc_pwr_decom}
\end{figure}

Fig.~\ref{fig:meas_adc_pwr_decom} illustrates the measured power composition of the prototype at a supply voltage of 0.9~V for sampling rates of 3.5~GS/s and 10.5~GS/s. Overall, the power consumption is dominated by the TDC. The clock and control signal generation circuitry accounts for 28–30\% of the total power consumption, which can potentially be amortized across interleaved channels. At 0.9~V, the measured power increases from 28.3~mW at 3.5~GS/s to 66.7~mW at 10.5~GS/s. The overall VTC is only approximately 4~mW at 10.5~GS/s, demonstrating high power efficiency. As the sampling rate increases by 3×, the total power increases by only 2.36×, leading to a reduction in energy per sample from 8.09~pJ/sample to 6.35~pJ/sample. This sublinear scaling is consistent with the proposed reset-free dual-edge operation, in which explicit reset-related switching is reduced and does not scale proportionally with sampling rate. 

The observed trend suggests that further improvements in overall energy efficiency at higher sampling rates are achievable through optimization of the TDC delay elements. Importantly, the power scaling behavior indicates that no new dominant power bottleneck emerges as sampling rate increases, supporting the suitability of the proposed reset-free dual-edge architecture for high-speed operation.

\subsection{Comparison with Prior Art}

Table~I compares the proposed ADC with state-of-the-art high-speed voltage-domain and time-domain designs. Prior ADCs typically achieve multi-GS/s operation through extensive TI or other techniques that introduce tradeoffs in architectural complexity, calibration overhead, and power consumption. While time-domain ADCs can generally enable higher single-channel sampling rates and smaller areas than their voltage-domain counterparts, many reported designs rely on folding VTCs or delay-tracking techniques, which further increase architectural complexity. In contrast, the proposed ADC introduces dual-edge reset-free quantization as a new architectural degree of freedom. By exploiting both rising and falling edges within a single conversion period, explicit reset phases in the VTC and TDC signal paths are eliminated, improving time utilization and relaxing the resolution–speed tradeoff.

From an area perspective, the proposed design achieves a core area of 0.0089~$\mathrm{mm^2}$, among the smallest reported for pipelined time-domain ADCs in this speed range. The compact area of the proposed design facilitates integration with TI by enabling a smaller input buffer array and a simplified clock distribution network. This reduces system-level overhead and makes the architecture well suited for next-generation high-speed wireline systems. While the measured SNDR, SFDR, and Walden FoM (859.25~pJ/step) are not as competitive as those of several prior works, the observed limitations are attributable to implementation-level factors rather than intrinsic architectural constraints, as discussed in Section~\ref{sec:meas}.

\section{Conclusion} \label{sec:concl}

This paper presented a dual-edge asynchronous pipelined SAR time-domain ADC that enables reset-free quantization by exploiting both rising and falling signal edges within a single conversion period. By eliminating explicit reset phases inherent to conventional single-edge time-domain architectures, the proposed approach expands effective waveform utilization and relaxes the fundamental resolution–speed tradeoff at high sampling rates.

A prototype fabricated in 22-nm FD-SOI validates the feasibility of dual-edge quantization in a fully asynchronous pipelined SAR framework. Measured results demonstrate operations of both rising- and falling-edge signal paths, and architectural scalability to higher sampling rates. The measured power consumption increases sublinearly with sampling rate, resulting in reduced energy per sample at higher speeds, which is consistent with the reset-free nature of the proposed architecture. The measured SNDR and linearity are currently limited by implementation factors, most notably stage-level delay mismatch in the TDC. The limitations do not reflect fundamental constraints of the dual-edge quantization concept and are expected to be addressed through improved delay tuning granularity in future revisions.

Overall, this work establishes dual-edge reset-free quantization as a viable and scalable architectural approach for high-speed time-domain ADCs, providing a new degree of freedom for addressing throughput, efficiency, and scalability challenges in next-generation wireline receivers.

\section{Acknowledgment} \label{sec:ackn}
The authors would like to thank the Natural Sciences and Engineering Research Council of Canada (NSERC), University of Toronto and Alphawave Semi. The authors also acknowledge GlobalFoundries and CMC Microsystems for chip fabrication and related services.

\bibliographystyle{IEEEtran}
\bibliography{ref}

\begin{IEEEbiography}
[{\includegraphics[width=1in,height=1.25in,clip,keepaspectratio]{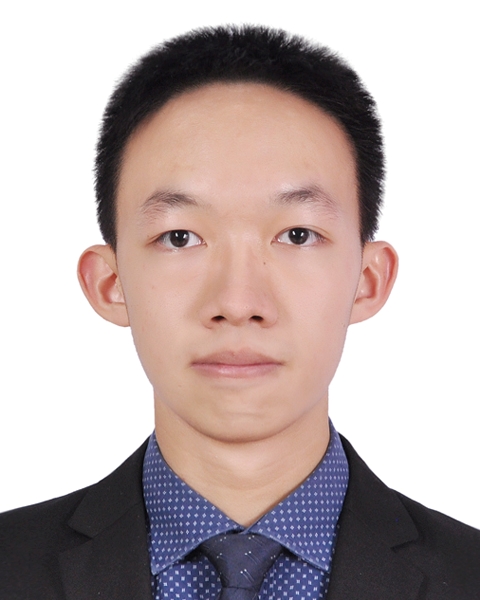}}]{Richard Zeng}(Graduate Student Member, IEEE) received the M.A.Sc. degree in the Edward S. Rogers Sr. Department of Electrical and Computer Engineering (ECE) at the University of Toronto in 2025, where he is currently pursuing the Ph.D. degree. His research interests include high-speed time-domain ADCs and AI-assisted CMOS RFICs. Since 2024, he has served as a conference volunteer at the IEEE International Solid-State Circuits Conference Saratoga Group, contributing to the technical and organizational execution of the conference. 
\end{IEEEbiography}

\begin{IEEEbiography}
[{\includegraphics[width=1in,height=1.25in,clip,keepaspectratio]{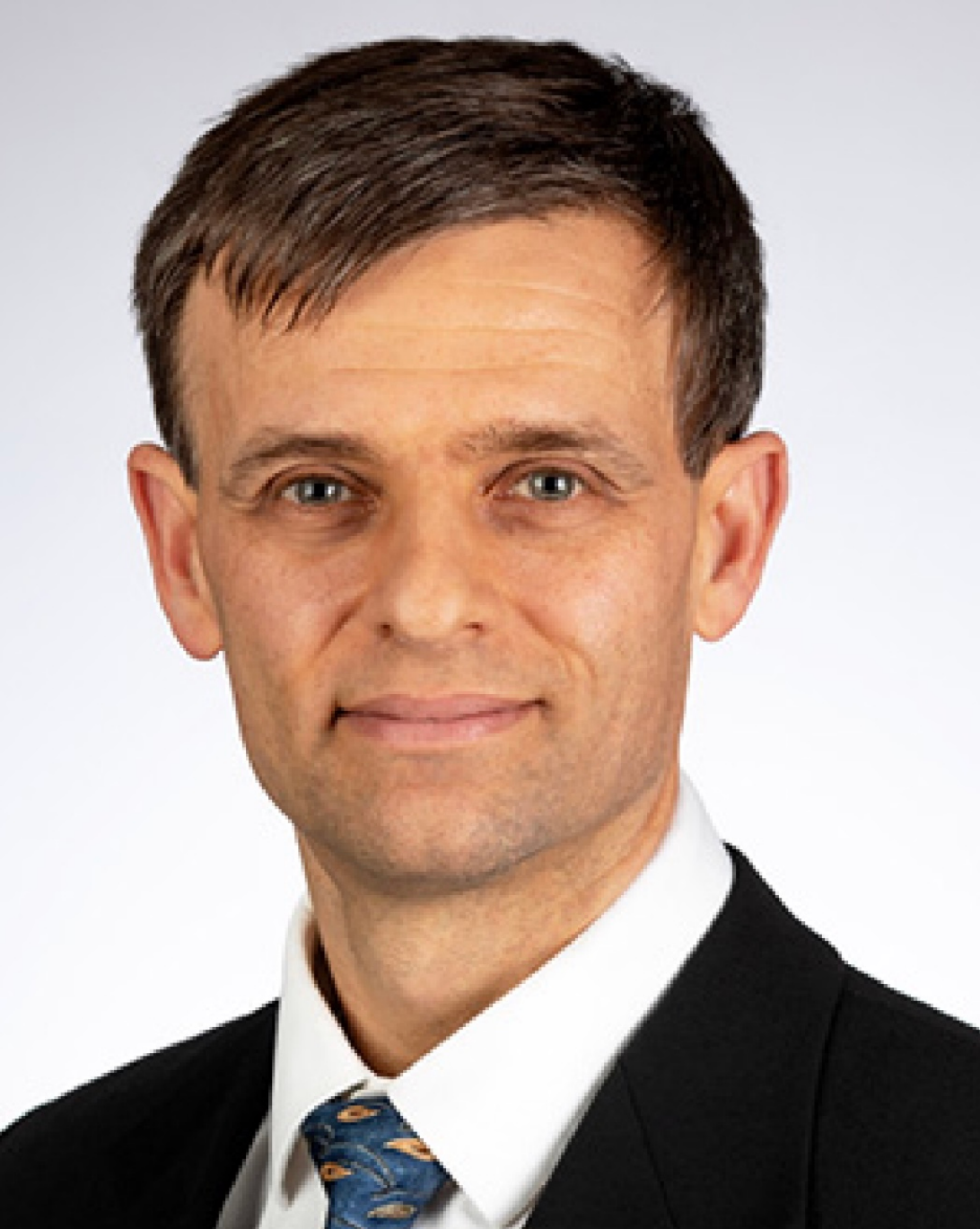}}]{Anthony Chan Carusone}(Fellow, IEEE) received the Ph.D. degree from the University of Toronto in 2002.

Since, he has been a Professor with Department of Electrical and Computer Engineering, University of Toronto. He has also been a consultant to industry in the areas of integrated circuit design and digital communication since 1997. He is currently the Chief Technology Officer with Alphawave Semi. He co-authored the popular textbooks Analog Integrated Circuit Design (along with D. Johns and K. Martin) and Microelectronic Circuits 8th edition (along with A. Sedra, K.C. Smith and V. Gaudet). He co-authored Best Student Papers at the 2007, 2008, 2011, and 2022 Custom Integrated Circuits Conferences, the Best Invited Paper at the 2010 Custom Integrated Circuits Conference, the Best Paper at the 2005 Compound Semiconductor Integrated Circuits Symposium, the Best Young Scientist Paper at the 2014 European Solid-State Circuits Conference, and Best Papers at DesignCon 2021 and 2023. He was the Editor-in-Chief of the \textsc{IEEE Transactions ON Circuits and Systems—II: Express Briefs} in 2009, an Associate Editor for the \textsc{IEEE Journal OF Solid-State Circuits} from 2010 to 2017, and the Editor-in-Chief of the \textsc{IEEE Solid-State Circuits Letters} from 2021 to 2023. He was a Distinguished Lecturer for the IEEE Solid-State Circuits Society from 2015 to 2017 and has served on the technical program committee of several IEEE conferences, including the International Solid-State Circuits Conference from 2016 to 2021.
\end{IEEEbiography}

\begin{IEEEbiography}
[{\includegraphics[width=1in,height=1.25in,clip,keepaspectratio]{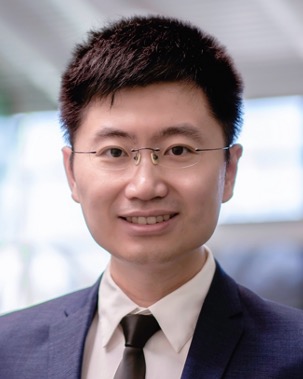}}]{Xilin Liu}(Senior Member, IEEE) obtained his Ph.D. degree from the University of Pennsylvania, Philadelphia, PA, USA, in 2017. He is currently an Assistant Professor in the Edward S. Rogers Sr. Department of Electrical and Computer Engineering (ECE) at the University of Toronto, Toronto, ON, Canada. His research interests include analog and mixed-signal IC design for low-power sensors, biomedical systems, and digital communications. He held industrial positions at Qualcomm Inc., San Diego, CA, USA between 2017 and 2021. 

Dr. Liu currently serves as an Associate Editor of the \textsc{IEEE Transactions of Biomedical Circuits and Systems (TBioCAS)} and the \textsc{IEEE Transactions on Circuits and Systems II: Express Briefs (TCAS-II)}. He also served on the committees of several CASS and SSCS conferences. 
\end{IEEEbiography}

\ifCLASSOPTIONcaptionsoff
  \newpage
\fi

\end{document}